\documentclass[usenatbib]{mn2e}

\usepackage[dvips]{graphicx}
\usepackage{amssymb,txfonts,color}
\usepackage[dvipdfm,colorlinks=true,linkcolor=blue,urlcolor=blue,citecolor=blue]{hyperref}

\newcommand{\apj}{{ApJ}}
\newcommand{\apjs}{{ApJS}}
\newcommand{\apjl}{{ApJL}}
\newcommand{\mnras}{{MNRAS}}
\newcommand{\aap}{{A\&A}}
\newcommand{\araa}{{ARA\&A}}

\def\be{\begin{equation}}
\def\ee{\end{equation}}

\newcommand{\msun}{{\rm M}_{\sun}}
\newcommand{\rxte}{{\it RXTE}}
\newcommand{\rxt}{{\it Rossi X-Ray Timing Explorer}}
\newcommand{\rosat}{{\it ROSAT}}
\newcommand{\cha}{{\it Chandra}}
\newcommand{\asca}{{\it ASCA}}
\newcommand{\xmm}{{\it XMM-Newton}}
\newcommand{\hb}{${\rm H\beta}\ $}
\newcommand{\mbh}{$M_{\rm BH}\ $}

\newcommand{\lxedd}{L_{\rm X}/L_{\rm Edd}}
\newcommand{\medd}{\dot{M}_{\rm Edd}}

\newbox\grsign \setbox\grsign=\hbox{$>$} \newdimen\grdimen \grdimen=\ht\grsign
\newbox\simlessbox \newbox\simgreatbox \newbox\simpropbox
\setbox\simgreatbox=\hbox{\raise.5ex\hbox{$>$}\llap
     {\lower.5ex\hbox{$\sim$}}}\ht1=\grdimen\dp1=0pt
\setbox\simlessbox=\hbox{\raise.5ex\hbox{$<$}\llap
     {\lower.5ex\hbox{$\sim$}}}\ht2=\grdimen\dp2=0pt
\setbox\simpropbox=\hbox{\raise.5ex\hbox{$\propto$}\llap
     {\lower.5ex\hbox{$\sim$}}}\ht2=\grdimen\dp2=0pt
\def\ga{\mathrel{\copy\simgreatbox}}
\def\la{\mathrel{\copy\simlessbox}}

\title[$\Gamma$-$\lxedd$ correlation in BH accretion systems]{Correlation between the photon index and X-ray luminosity of black hole X-ray binaries and active galactic nuclei: observations and interpretation}
\author[Q. -X. Yang et al.]
{Qi-Xiang Yang,$^{1,2}$ Fu-Guo Xie,$^{1}$\thanks{E-mail: fgxie@shao.ac.cn (FGX)} Feng Yuan,$^{1}$ Andrzej A. Zdziarski,$^{3}$ Marek Gierli\'nski,$^{4}$
\newauthor
Luis C. Ho$^{5,6,7}$ and Zhaolong Yu$^{1}$\\
\\
$^{1}$Key Laboratory for Research in Galaxies and Cosmology, Shanghai Astronomical Observatory,Chinese Academy of Sciences, 80 Nandan Road, Shanghai \\ 200030, China; \textcolor[rgb]{0.00,0.00,1.00}{\mbox{qxyang@shao.ac.cn}}; \textcolor[rgb]{0.00,0.00,1.00}{\mbox{fyuan@shao.ac.cn}}\\
$^{2}$University of Chinese Academy of Sciences, 19A Yuquan Road, Beijing 100049, China\\
$^{3}$Centrum Astronomiczne im.\ M. Kopernika, Bartycka 18, PL-00-716 Warszawa, Poland; \textcolor[rgb]{0.00,0.00,1.00}{\mbox{aaz@camk.edu.pl}}\\
$^{4}$Department of Physics, University of Durham, South Road, Durham DH1 3LE, UK; \textcolor[rgb]{0.00,0.00,1.00}{\mbox{marek.gierlinski@gmail.com}}\\
$^{5}$Kavli Institute for Astronomy and Astrophysics, Peking University, Beijing 100871, China \\
$^{6}$Department of Astronomy, Peking University, Beijing 100871, China\\
$^{7}$The Observatories of the Carnegie Institution for Science, 813 Santa Barbara Street, Pasadena, CA 91101, USA; \textcolor[rgb]{0.00,0.00,1.00}{\mbox{lho@obs.carnegiescience.edu}}
}

\begin{document}
\date{}
\pagerange{\pageref{firstpage}--\pageref{lastpage}}

\maketitle

\label{firstpage}

\begin{abstract}
We investigate the observed correlation between the 2--10 keV X-ray luminosity (in unit of the Eddington luminosity; $l_{\rm X} \equiv \lxedd$) and the photon index ($\Gamma$) of the X-ray spectrum for both black hole X-ray binaries (BHBs) and active galactic nuclei (AGNs). We construct a large sample, with $10^{-9}\la l_{\rm X}\la 10^{-1}$. We find that $\Gamma$ is positively and negatively correlated with $l_{\rm X}$ when $l_{\rm X}\ga 10^{-3}$ and $10^{-6.5}\la  l_{\rm X}\la  10^{-3}$ respectively, while $\Gamma$ is nearly a constant when $l_{\rm X}\la  10^{-6.5}$. We explain the above correlation in the framework of a coupled hot accretion flow -- jet model. The radio emission always come from the jet while the X-ray emission comes from the accretion flow and jet when $l_{\rm X}$ is above and below $10^{-6.5}$, respectively. More specifically, we assume that with the increase of mass accretion rate, the hot accretion flow develops into a clumpy and further a disc -- corona two-phase structure because of thermal instability. We argue that such kind of two-phase accretion flow can explain the observed positive correlation.
\end{abstract}
\begin{keywords}
accretion, accretion discs --- X-rays: general --- X-rays: binaries ---galaxies: active --- quasars: general.
\end{keywords}

\section{Introduction}

The X-ray spectra of both black hole (BH) X-ray binaries (BHBs) and active galactic nuclei (AGNs) can be roughly described by a power-law form, i.e., the photon number density being $N(E)\propto E^{-\Gamma}$, where $E$ is the photon energy and $\Gamma$ is the photon index. The photon index has been found to correlate with the X-ray luminosity. In the case of BHBs, Yuan et al.\ (\citeyear{Yuan2007}) found based on the data of two sources that $\Gamma$ is negatively (positively) correlated with the 1--100 keV X-ray luminosity, for the luminosity below (above) a turning point of the luminosity at $\sim$1--2 per cent of the Eddington luminosity, $L_{\rm Edd}$. This work was further developed by including more sources and similar results were found (e.g., Wu \& Gu \citeyear{Wu2008}; Cao, Wu \& Dong \citeyear{CWD2014}).  The turning point is constrained to be around $L_{\rm X}(0.5$--$25\ {\rm keV})\sim 1\%\ L_{\rm Edd}$.

In the case of AGNs, a positive correlation between $\Gamma$ and the X-ray luminosity was found in luminous sources (e.g., Wang, Watarai \& Mineshige \citeyear{Wang2004}; Shemmer et al.\ \citeyear{Shemmer2006,Shemmer2008}; Risaliti, Young \& Elvis\ \citeyear{Risaliti2009}; Zhou \& Zhao \citeyear{Zhou2010}; Wu et al.\ \citeyear{Wu2012}; Fanali et al.\ \citeyear{Fanali2013}; Brightman et al.\ \citeyear{Brightman2013}). On the other hand, a negative correlation was found in low-luminosity AGNs (hereafter LLAGNs; e.g., Gu \& Cao \citeyear{Gu2009}; Younes et al.\ \citeyear{Younes2011}; Hern\'{a}ndez-Garc\'{i}a et al.\ \citeyear{Hern2013}; Jang et al.\ \citeyear{Jang2014}). Combining both bright and dim AGNs, Constantin et al.\ (\citeyear{Constantin2009}, see also Veledina, Vurm \& Poutanen \citeyear{Vel2011}; Trichas et al.\ \citeyear{Trichas2013}) found positive and negative correlations similar to those of BHBs. The turning point was again found to be $L_{\rm bol}\sim 1\%L_{\rm Edd}$.

In all of the above works, the luminosities of the sources are actually not very low. For the case of extremely low-luminosity sources, Corbel et al.\ (\citeyear{Corbel2006}) studied the \cha\ observations of two BHBs (XTE J1550--564 and H1743--322) in their quiescent states and found a saturation of $\Gamma$, $\Gamma\sim 2.2$. Similarly, Plotkin, Gallo \& Jonker\ (\citeyear{Plotkin2013}) studied three BHBs (H1743--322, MAXI J1659--152, and XTE J1752--223) in their quiescent states, and found $\Gamma$ plateaus around $\Gamma\sim 2.1$ (also see Reynolds et al.\ \citeyear{Reynolds2014}). In the case of LLAGNs, the saturation of $\Gamma$ at $\sim 2.1$--2.2 has also been found (e.g., Wrobel et al.\ \citeyear{Wrobel2008}; G{\"u}ltekin et al.\ \citeyear{Gultekin2012}).

The correlations have been interpreted in various scenarios in literature. The first type of works is based on the truncated thin disc plus hot accretion flow model (hereafter abbreviated as the truncated disc model), see, e.g., Veledina, Vurm \& Poutanen\ (\citeyear{Vel2011}), Sobolewska et al.\ (\citeyear{Sobolewska2011}) and Gardner \& Done (\citeyear{Gardner2013}). In this model, the thin disc is truncated at a radius which decreases with the increasing accretion rate. Inside the truncation radius there is a hot accretion flow, which may also partially overlap with the disc. This scenario has been widely adopted in literature as it explains many observational aspects of both BHBs and LLAGNs (see a recent review by Yuan \& Narayan \citeyear{Yuan2014}). In this model, the X-ray radiation is from thermal Comptonization in the hot accretion flow of some soft seed photons (e.g., Zdziarski \citeyear{Zdziarski1998}; see review by Zdziarski \& Gierli{\'n}ski \citeyear{Zdziarski2004}; Done, Gierlinski \& Kubota\ \citeyear{Done2007}).

The seed photons can be either the blackbody emitted by the outer disc or synchrotron radiation emitted by the hot flow. In the former case, the flux of soft photons incident on the hot flow increase with increasing the accretion rate, as the outer disc moves closer to the black hole and the overlap with the hot flow increases. The increasing incident seed photon flux increases the cooling of the hot flow and thus steepens the X-ray spectrum, which results in a positive index-flux correlation (Zdziarski, Lubi\'nski \& Smith \citeyear{zls99}).
In the latter case, the increasing accretion rate increases the optical depth of the hot flow, which in turn increases the degree of synchrotron self-absorption. This reduces the soft photon luminosity with respect to the power dissipated in the hot flow, which then hardens the X-ray spectrum, resulting in a negative index-flux correlation. This indeed has been found in many hot-flow models (e.g., Esin, McClintock \& Narayan \citeyear{Esin1997}; Veledina et al.\ \citeyear{Vel2011}; Sobolewska et al.\ \citeyear{Sobolewska2011}). Veledina et al.\ (\citeyear{Vel2011}; see also Sobolewska et al.\ \citeyear{Sobolewska2011} and Gardner \& Done \citeyear{Gardner2013}) propose that the observed change of the correlation sign is due to the above.

Note that in the specific model of Veledina et al.\ (\citeyear{Vel2011}), see also Poutanen \& Veledina\ (\citeyear{Poutanen2014}) for a review, the bulk of synchrotron emission is produced by non-thermal rather than thermal electrons. Such electrons might exist in hot accretion flows (Malzac \& Belmont\ \citeyear{MB2009}; Poutanen \& Vurm\ \citeyear{Poutanen2009}; Nied{\'z}wiecki, Xie \& Stepnik \citeyear{Nied2014}; Nied{\'z}wiecki, Stepnik \& Xie \citeyear{Nied2015}) and then are likely to dominate the synchrotron emission (Wardzi{\'n}ski \& Zdziarski\ \citeyear{WZ01}).

Interestingly, in the related evaporated-corona model the cold disc within certain radius is totally evaporated into a corona, which thus represents a hot flow. Such corona is indeed dynamically, geometrically and radiatively very similar to an ADAF, and its presence can also explain the negative correlation (Qiao \& Liu \citeyear{Qiao2013a}; Qiao et al.\ \citeyear{Qiao2013b}). Moreover, at high accretion rates the innermost region of the corona collapses to form an inner cold disc close to the black hole (Liu, Meyer \& Meyer-Hofmeister \citeyear{Liu2006}). Seed photons from this newly formed disc will increase the cooling of the corona, which will then lead to a positive correlation (Qiao \& Liu \citeyear{Qiao2013a}; Qiao et al.\ \citeyear{Qiao2013b}). We note that the inner cold disc here play a very similar role to the cold clumps in our clumpy two-phase model, cf.\ Section \ref{theory}.

A competing scenario for the X-ray production is the standard disc-corona model (Liang \& Price \citeyear{Liang1977}; Galeev, Rosner \& Vaiana \citeyear{Galeev1979}; Haardt \& Maraschi \citeyear{Haardt1991}; Stern et al.\ \citeyear{S95}; Merloni \& Fabian \citeyear{Merloni2001}; Liu, Mineshige \& Ohsuga \citeyear{Liu2003}; Schnittman, Krolik \& Noble \citeyear{Schnittman2013}), where the thin disc extends all the way down to the radius of the innermost stable circular orbit (ISCO). In this model, the hot corona above the standard thin disc is responsible for the X-ray emission. The disc-corona can naturally explain the positive index-flux correlation (e.g. Cao \citeyear{Cao2009}; Cao, Wu \& Dong \citeyear{CWD2014}; Cao \& Wang \citeyear{Cao2014}), because the fractional dissipation in the corona decreases as the sources brightens. We note that such interpretation has several shortages. First, generally disc-corona model can only reproduce a spectrum typical for the soft state (maybe also soft intermediate state) of BHBs, where the thermal emission from the cold disc is strong. Second, it can only explain the positive correlation with $\Gamma\ga 2$ (Haardt, Maraschi \& Ghisellini\ \citeyear{HMG94}; Stern et al.\ \citeyear{S95}). Third, it can not explain the observed negative correlation.

The presence of the correlation is likely to have implications for our understanding of accretion physics and radiative processes. In this paper, our first goal is to construct a sample which is larger than all previous ones and study the correlation between photon index and X-ray luminosity. Our sample includes both BHBs and AGNs and covers a wide range of luminosity, from extremely sub-Eddington to close to Eddington.

The second goal is to interpret the correlation  based on the truncated thin disc plus hot accretion flow and jet model (hereafter abbreviated as the accretion-jet model). Compared to the previous works based on similar model frameworks, our model is different in the following aspects:  (1) It is based on more precise calculations of the dynamics and radiation of the hot accretion flow. (2) The jet component is included, which plays an important role on the X-ray emission at low luminosities. (3) We invoke a new model -- a two-phase accretion flow -- to explain the correlation of high-luminosity sources. (4) The causes of the scatter of the correlation between individual sources are also studied.

The paper is organized as follows. In Section \ref{sample}, we show the correlation based on available observations. A theoretical interpretation is given in Sections \ref{theory}--\ref{results}. Finally, we summarize and discuss the results in Section\ \ref{summary}.

\section{The Sample}
\label{sample}

\begin{figure*}
\centering
\includegraphics[width=15.cm]{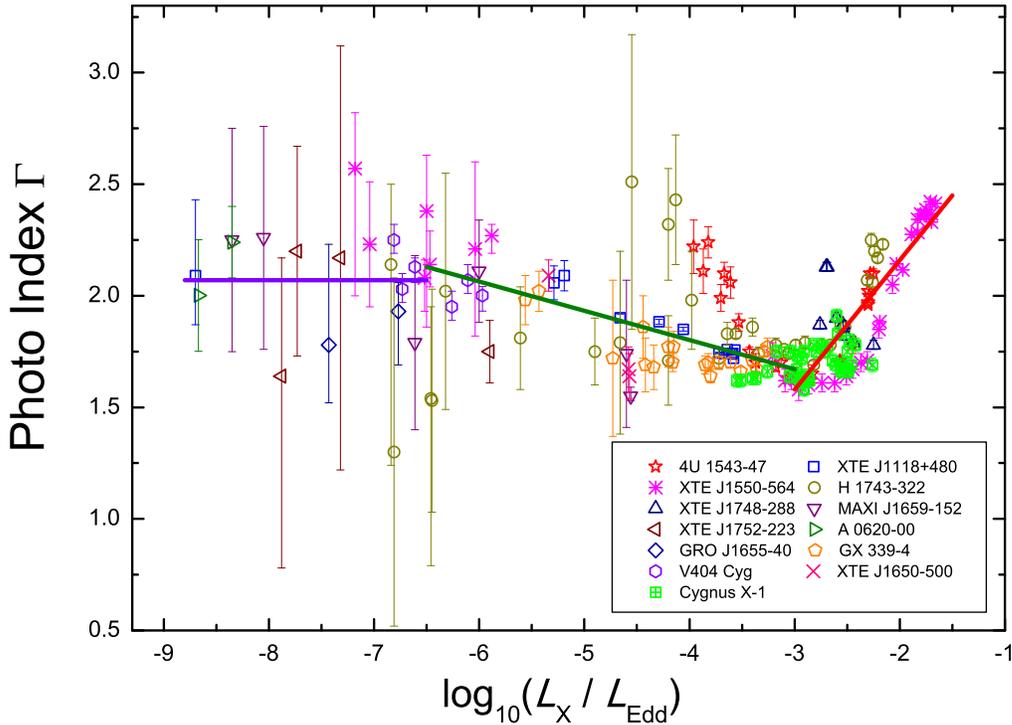}
\caption{The X-ray photon index, $\Gamma$, versus the 2--10 keV X-ray luminosity (in the Eddington unit; $\lxedd$) correlation for the BHB sample. The lines show the phenomenological fits of Section \ref{BHBs}.
}\label{fig-bhb-sample}
\end{figure*}

Our observational data are collected from literature. We select sources with measurements of their distance ($d$) and the black hole mass ($M_{\rm BH}$). We accept several BHBs which only have reasonable estimates on $M_{\rm BH}$ and/or $d$. We require that the sources have relatively good observations in the X-ray band so that we can obtain the photon index $\Gamma$ and the hard X-ray luminosity in the 2--10 keV band ($L_{\rm X}$ hereafter). For some sources which only have luminosities at a specified energy, e.g., 1 keV, or over different X-ray bands, e.g., 0.5--10 keV, we calculate their 2--10 keV luminosity using $\Gamma$. We exclude sources with extreme values of $\Gamma$ (following the suggestion by Risaliti, Young \& Elvis\ \citeyear{Risaliti2009}), namely $\Gamma > 3$ or $\Gamma < 1$, as those measurements usually bear large uncertainties. Our final sample includes 523 observational data points, among which 195 data points are from BHBs and 328 data points are from AGNs.

\subsection{The BHB sample}
\label{BHBs}

For the sample of  BHBs, we collect data for the quiescent, hard and intermediate states (cf.\ Remillard \& McClintock \citeyear{RM2006}, Done, Gierli\'nski \& Kubota \citeyear{Done2007} and Belloni \citeyear{Belloni2010} for the definition of spectral states in BHBs). The sample includes the data as below.
\begin{itemize}
\item 17 data points for GX 339--4 in its hard state during its outburst decay in 2011. The values of $\Gamma$ and the flux are from \rxt\ (\rxte) observations in Din\c{c}er et al.\ (\citeyear{Dincer2012}), where a multi-color disc blackbody component and a power-law component are used in the spectral fitting, taking Galactic absorption and broad iron lines (if present) into account. We assume $M_{\rm BH}=5.8\, \msun$ (Hynes et al.\ \citeyear{Hynes2003}) and $d=8$ kpc (Zdziarski et al.\ \citeyear{Zdziarski2004b}).
\item 41 data points for Cygnus X-1 in its hard and intermediate states. The values of $\Gamma$ and flux are taken from {\it Ginga\/} and \rxte\ observations given in Ibragimov et al.\ (\citeyear{Ibragimov2005}), where thermal Comptonization with reflection was used for the X-ray spectral fitting. The \mbh and $d$ are $\sim16\, \msun$ and 1.86 kpc, respectively, taken from Zi\'{o}{\l}kowski (\citeyear{Ziolkowski2014}) and Reid et al.\ (\citeyear{Reid2011}).
\item 23 data points for 4U 1543--47 in the hard and intermediate states during the decay of its 2002 outburst. The X-ray photon index and the flux are taken from \rxte\ observations (Kalemci et al.\ \citeyear{Kalemci2005}, with the same spectral fitting model as that adopted for GX 339--4), and $M_{\rm BH}=9.4\, \msun$ and $d=7.5$ kpc are from Park et al.\ (\citeyear{Park2004}).
\item 12 data points for XTE J1118+480 in its hard state. The values of $\Gamma$ and flux of XTE J1118+480 are from \rxte\ observations in Kalemci (\citeyear{Kalemci2002}, with the same spectral fitting model as that adopted for GX 339--4), and its black hole mass ($8.53\, \msun$) and distance (1.72 kpc) are from Gelino et al.\ (\citeyear{Gelino2006}).
\item 27 data points for XTE J1550--564 in its hard and intermediate states. The values of $\Gamma$ and flux are taken from \rxte\ observations in Kalemci et al.\ (\citeyear{Kalemci2004}; 14 data points), and Tomsick et al.\ (\citeyear{Tomsick2001}; 13 data points). The spectral fitting model is the same as that adopted for GX 339--4. The \mbh and $d$ are $9.1\, \msun$ and 4.38 kpc, respectively, from Orosz et al.\ (\citeyear{Orosz2011}).
\item 27 data points for H 1743--322 in its hard and intermediate states during the decay of its 2003 outburst. The values of $\Gamma$ and the flux are taken from \rxte\ observations in Kalemci et al.\ (\citeyear{Kalemci2006}, with the same spectral fitting model as that adopted for GX 339--4). The black hole mass and the distance are $10\, \msun$ and 8.5 kpc, respectively, from Steiner et al.\ (\citeyear{Steiner2012}).
\item 8 data points for XTE J1748--288 in the hard state during its 1998 outburst. The values of $\Gamma$ and the flux are from \rxte\ observations in Revnivtsev et al.\ (\citeyear{Revnivtsev2000}, the same spectral fitting model as GX 339--4). The black hole mass ($10\, \msun$) and distance (8 kpc) are from Wu \& Gu (\citeyear{Wu2008}).
\item 4 data points for V404 Cyg in its quiescent state from the long-term observations by \cha, \xmm\ and {\it Suzaku}. The values of $\Gamma$ and the flux are obtained from Reynolds et al.\ (\citeyear{Reynolds2014}, absorbed power-law model is adopted). The $M_{\rm BH}=12\, \msun$  and $d=2.39$ kpc are from Shahbaz et al.\ (\citeyear{Shahbaz1994}) and Miller-Jones et al.\ (\citeyear{Miller2009}), respectively.
\item 36 data points for 10 BHBs, observed by \cha\ as they fade into the quiescent state. These sources include A 0620-00, V404 Cyg, GRO J1655--40, GX 339--4, H 1743--322, MAXI J1659--152, XTE J1118+480, XTE J1550--564, XTE J1650--500 and XTE J1752--223. The values of $\Gamma$ and the luminosity are from Plotkin, Gallo \& Jonker (\citeyear{Plotkin2013}; and references therein), where a simple absorbed power-law model is adopted for the spectral fitting. Information of the black hole mass and the distance are also provided. We note that the disc emission component is expected to be absent at such low luminosities and there will be no difference when modelled like the case of GX 339--4.
\end{itemize}

Fig.\ \ref{fig-bhb-sample} shows the relationship between the photon index $\Gamma$ and the X-ray luminosity (in unit $L_{\rm Edd}$, $l_{\rm X}\equiv L_{\rm X}/L_{\rm Edd}$, which we hereafter call the luminosity ratio) for the BHB sample. The uncertainty of $l_{\rm X}$, which is at most $\sim 0.5$ dex, is not shown in the figure for the purpose of clarity. We see that $l_{\rm X}$ covers a wide range, from $\sim 10^{-8.5}$ to $\sim 10^{-1.5}$. The photon index is in the range of 1.5--2.5 for most of the sources. We note that all the faint observations with $L_{\rm X} \lesssim 10^{-5} L_{\rm Edd}$ shown in Fig.\ \ref{fig-bhb-sample} are observed by CCD detectors aboard on, e.g., \cha, \xmm\/ or {\it Suzaku\/} (Plotkin et al.\ \citeyear{Plotkin2013}; Reynolds et al.\ \citeyear{Reynolds2014}), which avoid the possible Galactic ridge emission effect as observed by non-imaging instruments like \rxte. Moreover, we note that all sources, except Cyg X-1, have been fitted with almost the same spectral model. Still, Cyg X-1 shows no clear offset with respect to other sources, see Fig.\ \ref{fig-bhb-sample}. Thus, this issue may affect our results at the range $10^{-3.5} L_{\rm Edd} \lesssim L_{\rm X} \lesssim 10^{-2.5} L_{\rm Edd}$ only in a minor way.

Roughly, we see in Fig.\ \ref{fig-bhb-sample} that the correlation consists of three branches, namely (1) $l_{\rm X} \ga  10^{-3}$, (2) $10^{-6.5} \la  l_{\rm X} \la  10^{-3}$ and (3) $l_{\rm X} \la  10^{-6.5}$. Using weighted least-$\chi^2$ linear fitting between $\Gamma$ and $\log_{10} (\lxedd)$, we have found the following relationships (the uncertainties below are indicated at 1-$\sigma$ level). For the high-luminosity branch ($l_{\rm X} \ga  10^{-3}$), $\Gamma$ and $l_{\rm X}$ are positively correlated,
\be \Gamma = (0.58\pm 0.01)\ \log_{10}(\lxedd) + (3.32\pm 0.02). \label{eq:pos-corr-BHB} \ee
For the moderate-luminosity branch ($10^{-6.5} \la  l_{\rm X} \la  10^{-3}$), an anti-correlation between $\Gamma$ and $l_{\rm X}$ is observed,
\be \Gamma = (-0.13\pm 0.01)\ \log_{10}(\lxedd) + (1.28\pm 0.02). \label{eq:anti-corr-BHB} \ee
The above V-shaped correlation between $\Gamma$ and $l_{\rm X}$ is consistent with the previous findings using more limited sample sizes (e.g. Yuan et al.\ \citeyear{Yuan2007}; Wu \& Gu \citeyear{Wu2008}; Sobolewska et al.\ \citeyear{Sobolewska2011}).

Finally, at the lowest luminosity branch ($l_{\rm X} \la  10^{-6.5}$), $\Gamma$ is nearly independent of $l_{\rm X}$ (the linear correlation slope is $\approx 0.02$). We thus use a constant to fit this part and found,
\be \langle \Gamma \rangle \approx 2.07\pm 0.06. \label{eq:flat-corr-BHB} \ee
The constant feature of $\Gamma$ in this branch is consistent with Plotkin, Gallo \& Jonker\ (\citeyear{Plotkin2013}; see also Reynolds et al.\ \citeyear{Reynolds2014}).

Theoretical understanding of these findings is presented in Sections \ref{theory} \& \ref{results}. We want to emphasize that the observed correlation has a relatively large scatter, especially in the moderate-luminosity branch. This is also the case of the AGN sample, as we will show in Section \ref{AGNs}. A major cause of the diversity is that different sources have different correlation slope, which we show in Section \ref{individualBHB}.

\subsection{The AGN sample}
\label{AGNs}

\begin{figure*}
\centering
\includegraphics[width=15.cm]{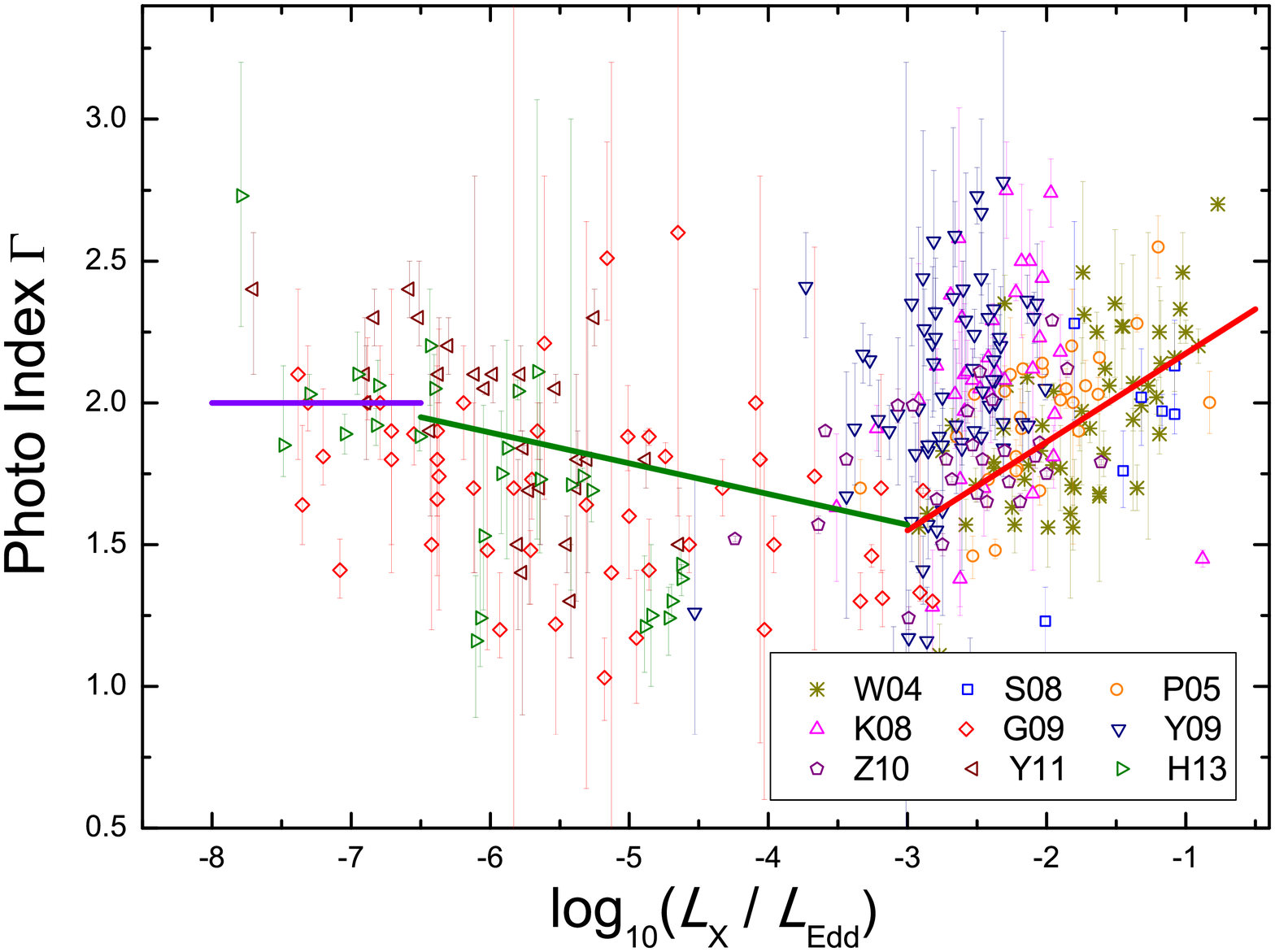}
\caption{The photon index $\Gamma$ versus the 2--10 keV X-ray luminosity correlation for the AGN sample.
\protect \\ The references: W04: Wang et al.\ \citeyear{Wang2004};  S08: Shemmer et al.\ \citeyear{Shemmer2008};  P05: Piconcelli et al.\ \citeyear{Piconcelli2005};  K08: Kelly et al.\ \citeyear{Kelly2008};  G09: Gu et al.\ \citeyear{Gu2009}; Y09: Young et al.\ \citeyear{Young2009};   Z10: Zhou et al.\ \citeyear{Zhou2010};  Y11: Younes et al.\ \citeyear{Younes2011};  H13: Hern\'{a}ndez-Garc\'{i}a et al.\ \citeyear{Hern2013}.}\label{fig-agn-sample}
\end{figure*}

We compile the AGN sample shown in Fig.\ \ref{fig-agn-sample}. For sources with only redshift information, we adopt a flat cosmological model with $H_{\rm 0}=70$ km s$^{-1}$ Mpc$^{-1}$ and $\Omega_{\Lambda}=0.73$ (Hinshaw et al.\ \citeyear{Hinshaw2013}) to derive their luminosity distances. The 2--10 keV energy range is also defined in the local rest frame of individual sources.

There are various ways to estimate the black hole mass (see Kormendy \& Ho \citeyear{KH2013} for a recent review), and the reverberation mapping method is among the most reliable ones. For the reverberation technique, depending on redshift, different broad emission lines are used, e.g., H$\beta$, Mg{\sc ii} and C{\sc iv}. As discussed in Risaliti, Young \& Elvis\ (\citeyear{Risaliti2009}), Mg{\sc ii} is a doublet and it also suffers the contamination of Fe{\sc ii} emission. Also, C{\sc iv} is a high-ionization line, related to the inner broad line regions where the gas may have non-virialized components. For sources whose black hole mass are derived through reverberation-mapping method, we limit ourselves to those using \hb line (cf.\ Section \ref{sec:caveats} for a discussion, especially on sources using C{\sc iv} line). Our AGN sample contains the sources as listed below.
\begin{itemize}
\item 35 quasars from Kelly et al.\ (\citeyear{Kelly2008}), with only sources whose $M_{\rm BH}$ are estimated by the \hb reverberation method selected. These sources have \rosat\ and/or \cha\ observations. The values of $\Gamma$ and $L_{\rm X}$ are from Kelly et al.\ (\citeyear{Kelly2007}), where an absorbed power-law model is adopted for the spectral fitting.

\item 68 quasars selected from the {\it SDSS}/\xmm\ quasar survey by Young et al.\ (\citeyear{Young2009}), where the absorbed power-law model is applied for the spectral fitting. Again, we only select the sources whose $M_{\rm BH}$ are estimated through the \hb reverberation method.

\item 7 luminous AGNs from Shemmer et al.\ (\citeyear{Shemmer2006,Shemmer2008}). The X-ray spectral fitting is through absorbed power-law model.

\item 56 AGNs consisting of 29 narrow-line Seyfert 1s and 27 broad-line Seyfert 1s from Wang, Watarai \& Mineshige\ (\citeyear{Wang2004}). The values of $\Gamma$ and $L_{\rm X}$ are from X-ray observations by {\it ASCA}, where absorbed power-law + iron line emission model is applied for the spectral fitting. Various methods are applied to derive their black hole masses.

\item 51 LLAGNs consisted of 25 LINERs and 26 local Seyferts, observed by  \xmm, \cha, and \asca, as assembled from literature by Gu \& Cao (\citeyear{Gu2009}). Various methods are adopted for the X-ray spectral fitting.

\item 29 Palomar Green quasars from Piconcelli et al.\ (\citeyear{Piconcelli2005}). Absorbed power-law model is adopted for for the X-ray spectral fitting. The \mbh of these sources are obtained from Vestergaard \& Peterson (\citeyear{Vestergaard2006}) and Inoue et al.\ (\citeyear{Inoue2007}).

\item 25 radio-quiet AGNs with 27 data points, observed by \xmm, from Zhou \& Zhao (\citeyear{Zhou2010}). Absorbed power-law + accretion disc emission model is adopted for the X-ray spectral fitting.

\item 13 LINERs with 28 data points, observed by \xmm\ and \cha, from Younes et al.\ (\citeyear{Younes2011}). Absorbed power-law + thermal emission model is adopted for the X-ray spectral fitting.

\item 6 LINERs with 27 data points, observed by \xmm\ and \cha, from Hern\'{a}ndez-Garc\'{i}a et al.\ (\citeyear{Hern2013}). The \mbh data are from Gonz\'alez-Mart\'in et al.\ (\citeyear{Gonz2009}). Various fitting models of the X-ray emission are adopted, i.e., (absorbed) power-law component plus possibly a thermal emission component, or two power-laws with the same slope (cf.\ their paper for fitting details.).
\end{itemize}

For the majority of our AGN sample, the X-ray spectral fitting used a power-law with Galactic bound-free absorption. We also find that the data from different works are roughly overlap with each other on the $\Gamma$-$l_{\rm X}$ plane. We thus do not expect any significant systematic bias due to different spectral fitting techniques. Fig.\ \ref{fig-agn-sample} shows the correlation between $\Gamma$ and $l_{\rm X}$ for the AGN sample. The luminosity ratio $l_{\rm X}$ of our AGN sample covers a wide range, from $\sim 10^{-8}$ to $\sim 10^{-0.5}$. Taking the same $l_{\rm X}$ boundaries and fitting method as those of the BHB sample, we find that for the luminous branch ($l_{\rm X}\ga  10^{-3}$), the correlation is
\be \Gamma = (0.31\pm 0.01)\ \log_{10}(\lxedd) + (2.48\pm 0.02).\label{eq:pos-corr-agn} \ee
For the moderate luminosity branch ($10^{-6.5}\la  l_{\rm X} \la  10^{-3}$), a negative correlation is found,
\be \Gamma = (-0.10 \pm 0.02)\ \log_{10}(\lxedd) + (1.27\pm 0.03). \label{eq:anti-corr-agn} \ee
The above two branches are consistent with previous works that focus on luminous AGNs (e.g., Shemmer et al., \citeyear{Shemmer2008}; Zhou \& Zhao\ \citeyear{Zhou2010}; Fanali et al.\ \citeyear{Fanali2013}; Brightman et al.\ \citeyear{Brightman2013}) or LLAGNs (e.g., Gu \& Cao \citeyear{Gu2009}; Younes et al.\ \citeyear{Younes2011}; Hern\'{a}ndez-Garc\'{i}a et al.\ \citeyear{Hern2013}; Jang et al.\ \citeyear{Jang2014}). Finally, the photon index $\Gamma$ is roughly a constant for $l_{\rm X}\la  10^{-6.5}$,
\be \Gamma \approx 2.\ee
However, this result is not as solid as that of the BHB sample because of the fewer observational points.

Theoretical explanation of AGNs is provided in Sections \ref{theory} \& \ref{results}. For a given luminosity ratio, the scatter in $\Gamma$ of the AGN sample is much larger compared to that of the BHB sample. As in the case of BHBs, we later examine in Section \ref{individualAGN} the negative correlation between $\Gamma$ and $l_{\rm X}$ for one individual source, NGC~7213, which has a relatively good coverage in $l_{\rm x}$ (Emmanoulopoulos et al.\ \citeyear{Emmanoulopoulos2012}).

\section{The theoretical model}
\label{theory}

The correlations described above can provide important constrains on theoretical models for BHBs and AGNs. In this paper we investigate whether they can be explained in the framework of the coupled hot accretion flow -- jet model (e.g., Yuan, Cui \& Narayan \citeyear{YCN2005}), which is developed from previous works by Narayan (\citeyear{Narayan1996}), Esin, McClintock \& Narayan (\citeyear{Esin1997}), and Poutanen, Krolik \& Ryde (\citeyear{PKR1997}). This model has been applied to the quiescent and hard states of BHBs and the LLAGNs in general (see the review by Yuan \& Narayan \citeyear{Yuan2014}).

When the accretion rate is high, the X-ray emission will be dominated by the radiation from the hot accretion flow and that from the jet can be neglected (see also Zdziarski et al.\ \citeyear{Zdziarski2014}). With the decrease of the accretion rate, however, the X-ray emission from ADAF ($\propto\dot{M}^{2 - 3}$) decreases faster than that from jet ($\propto\dot{M}_{\rm jet}\propto\dot{M}$, where $\dot{M}_{\rm jet}$ is the mass loss rate in the jet.). Therefore, when the luminosity is below a critical value $L_{\rm crit}$, we {\it predict} the X-ray emission will be dominated by the synchrotron emission from the jet (Yuan \& Cui \citeyear{Yuan2005}). This prediction has been confirmed for observations of the faint LLAGNs and the quiescent state of BHBs (e.g., Wu, Yuan \& Cao\ \citeyear{Wu2007}; Pellegrini et al.\ \citeyear{Pellegrini2007}; Wrobel et al.\ \citeyear{Wrobel2008}; Pszota et al.\ \citeyear{Pszota2008}; Yuan, Yu \& Ho\ \citeyear{YYH2009}; de Gasperin et al.\ \citeyear{deGasperin2011}; Younes et al.\ \citeyear{Younes2012}; Xie, Yang \& Ma \citeyear{Xie2014}).

In addition to the spectrum, the correlation between the radio and X-ray emission observed in BHBs and AGNs provides an additional constraint (Corbel et al.\ \citeyear{Corbel2003}; Corbel et al.\ \citeyear{Corbel2013}; Merloni, Heinz \& di Matteo\ \citeyear{Merloni2003}; Falcke, K\"{o}rding \& Markoff\ \citeyear{Falcke2004}; G\"ultekin et al.\ \citeyear{Gultekin2009}). Yuan \& Cui (\citeyear{Yuan2005}) show that the correlation can be well explained by the coupled hot accretion -- jet model. Moreover, this model {\it predicts} that the correlation should steepen when the luminosity of the system is lower than $L_{\rm crit}/L_{\rm Edd}$, because then both the radio and X-ray radiation will be dominated by the jet. This prediction is well confirmed by observations of low-luminosity AGNs (e.g., Wrobel et al.\ \citeyear{Wrobel2008}; Yuan et al.\ \citeyear{YYH2009}; de Gasperin et al.\ \citeyear{deGasperin2011}; Younes et al.\ \citeyear{Younes2012}). However, three quiescent BHBs currently available (XTE J1118+480 -- Gallo et al.\ \citeyear{Gallo2014}; A 0620-00 -- Gallo et al.\ \citeyear{Gallo2006}; and V404 Cyg -- Corbel, Koerding \& Kaaret\ \citeyear{Corbel2008}) apparently do not confirm the prediction. On one hand, this is perhaps not so surprising given the statistical sense of the correlation. On the other hand, we note that this observational result appears to be not robust. Namely, there is no hard state data for A0620-00, the observation of V404 Cyg appears to be not dim enough to show the predicted steepening, and the radio detection to XTE J1118+480 is only $3\sigma$.  In conclusion, further investigations of the nature of the quiescent state of BHBs are highly desired.

Below, we describe in some detail the hot accretion flow and the jet model components.

\subsection{Hot Accretion Flow}
\label{hotaccretion}

The hot accretion flows include the advection-dominated accretion flow (ADAF;  Narayan \& Yi \citeyear{Narayan1994,Narayan1995}) and the luminous hot accretion flow (LHAF; Yuan \citeyear{Yuan2001,Yuan2003}) at low and high accretion rate regimes, respectively. We refer the reader to Yuan \& Narayan (\citeyear{Yuan2014}) for a detailed description on the related equations, dynamical structure and radiative properties of hot accretion flows. We provide only a brief review below.

In ADAFs, most of the energy released from gravitational energy is stored as thermal energy of the gas and advected inwards, thus the flow will become hot and tenuous. Correspondingly, synchrotron, bremsstrahlung and Compton scattering are the main radiative processes. As the accretion rate increases (and the system becomes brighter), the advection term becomes less important, and it even becomes negative as the flow enters the LHAF regime. The LHAF is thermally unstable (Yuan \citeyear{Yuan2003}). However, the hot flow can remain relatively stable if the accretion rate is not too high because then the accretion time-scale is shorter than the growth time-scale of thermal instability. We label this as Type I LHAF. When the accretion rate is higher than some critical value, the growth time-scale of the thermal instability is shorter than the accretion time-scale. Consequently, cold dense clumps may be formed in the hot gas (Yuan \citeyear{Yuan2003}; for similar ideas, see e.g. Shapiro, Lightman \& Eardley\ \citeyear{SLE76}; Guilbert \& Rees\ \citeyear{GR1988}; Ferland \& Rees\ \citeyear{FR1988}; Celotti, Fabian \& Rees\ \citeyear{CFR1992}; Collin-Souffrin et al.\ \citeyear{CS1996}; Kuncic, Celotti \& Rees\ \citeyear{KCR1997}; Krolik\ \citeyear{Krolik1998}). We call this a two-phase accretion flow or Type II LHAF. Finally, the flow may evolve into a disc-corona structure when the accretion rate is high enough (see Section \ref{summary} for more discussions).

ADAFs and Type I LHAFs, with their dynamics and the radiation well developed (Yuan \& Narayan \citeyear{Yuan2014}), are responsible for the X-ray emission of low and moderate luminosity sources, where the X-ray comes from the Comptonization of synchrotron emission in the hot flow. Two-phase accretion flows (Type II LHAFs), on the other hand, can explain the X-ray emission of luminous sources, as suggested by Yuan \& Zdziarski (\citeyear{Yuan2004}). In two-phase accretion flows, we expect that, as the total (cold and hot) accretion rate increases, the instability grows and a higher fraction of the accreting material will be in the cold phase. Consequently, the thermal emission from cold clumps will become more and more important compared to the synchrotron emission from hot gases, and it will become the dominant seed photon source. Due to the complexity of the coupling between the cold and hot phases, we follow Yuan \& Zdziarski (\citeyear{Yuan2004}) to simplify the calculation of the dynamics of two-phase accretion flow by replacing the electron energy equation with the Compton $y$-parameter,
\be y = (4\theta_{\rm e}+16\theta_{\rm e}^2)(\tau+\tau^2), \label{eq:ypar} \ee
where $\theta_{\rm e} = kT_{\rm e}/m_{\rm e} c^2$, $T_{\rm e}$ is the electron temperature and $\tau$ is the optical depth in the direction perpendicular to the disc. Moreover, the Compton $y$-parameter is expected to be anti-correlated with the accretion rate (and the X-ray luminosity; see Section \ref{explanation} for further discussion on this))\footnote{In reality, the Compton $y$-parameter will vary with the radius, here we assume it a constant for simplicity.}. Since our two-phase accretion model is still very crude and the $y$ -- $\dot{M}$ relationship cannot be determined from first principles, we arbitrarily assume,
\be y\approx0.7 \left({\dot{M}(R_{\rm S})\over \medd}\right)^{-3/4},\label{eq:yform}\ee
where $\medd = L_{\rm Edd}/c^2$ is the Eddington accretion rate and $R_{\rm S}$ is the Schwarzschild radius. Obviously the slope of the positive correlation (cf.\ the red stars in Figs.\ \ref{fig-bhb-theory} \& \ref{fig-agn-theory}) will be affected by this assumption.

The total luminosity of the two-phase accretion flow can then be obtained from the power input to the electrons (viscous heating to electrons and Coulomb electron-ion energy exchange; for details see Xie \& Yuan \citeyear{Xie2012}.). We calculate the photon index $\Gamma$ of the X-ray spectrum through the formulae $\Gamma \approx 2.25 y^{-2/9}$ (Beloborodov \citeyear{Belo1999}). We use the total luminosity and $\Gamma$ to derive the X-ray luminosity at a given range of frequency, assuming a simple power-law spectrum with exponential cutoffs at the high and low ends, i.e., $L_E\sim E^{1-\Gamma} \exp(-E_{\rm min}/E)\exp(-E/E_{\rm max})$. The value of $E_{\rm max}$ is determined by the electron temperature (as $E_{\rm max} = k T_{\rm e}$) at the radius where the emission peaks. We fix $E_{\rm min}=0.02~{\rm keV}$, which roughly corresponds to the temperature of the cold clumps. We note that the temperature of the cold clumps at a given Eddington ratio will likely be higher in BHBs than in AGNs, $\propto M_{\rm BH}^{-1/4}$ if the clump sizes scale with $R_{\rm S}$. Our results will be sensitive to the value of $E_{\rm min}$ if $\Gamma > 2$. Since we actually only have one calculation for each (BHB and AGN) subsample (cf. Section \ref{general-results} and Figs.\ \ref{fig-bhb-theory}\--\ref{fig-agn-theory}), we adopt for simplicity the same $E_{\rm min}$ for both BHBs and AGNs.

We note that because of the existence of cold clumps in the innermost region of the accretion flow, the two-phase accretion flow scenario may also be able to explain the possible existence of the dim thermal component and the broad iron $K\alpha$ line in the BHB hard state (e.g., Reis, Fabian \& Miller\ \citeyear{Reis2010}) with no need for a standard thin disc to extend down to the ISCO (see Liu et al.\ \citeyear{Liu2007,Liu2011} and Taam et al.\ \citeyear{Taam2008} for a similar idea; but see Nied{\'z}wiecki, Xie \& Stepnik \citeyear{Nied2014} for the attribution of the thermal component as the small bump in the curved Comptonized spectrum of ADAF.). However, we have not performed here any quantitative calculations of this effect. On the other hand, the optical/ultraviolet disc emission is strong and may dominate the bolometric luminosity even when the X-ray spectrum is hard, e.g., in the case of NGC~4151 (Lubi{\'n}ski et al. \citeyear{lubinski10}). We stress we have not taken such components into account in our modeling.

The main parameters of hot accretion flow include the outflow parameter $s$ (in the form $\dot{M}(R) \propto R^s$. $R$ is the radial location), the viscous parameter $\alpha$, the magnetic parameter $\beta$, which describes the ratio of the gas to magnetic pressure, and the electron heating parameter $\delta$, which describes the fraction of the turbulent dissipation which directly heats electrons. These parameters have strong effects on the radiative efficiency of hot accretion flow (Xie \& Yuan \citeyear{Xie2012}). Numerical simulations show that if the viscosity $\alpha$ is due to the magnetic stress associated with the magneto-hydrodynamical (MHD) turbulence driven by the magneto-rotational instability (Balbus \& Hawley \citeyear{Balbus1991,Balbus1998}), then the values of $\alpha$ and $\beta$ are not independent (Blackman, Penna \& Varniere\ \citeyear{Blackman2008}). Still, we treat here $\alpha$ and $\beta$ as two independent parameters. The physical mechanisms associated with $\delta$ include magnetic reconnection (Bisnovatyi-Kogan \& Lovelace \citeyear{BKL1997}; Quataert \& Gruzinov \citeyear{Quataert1999}; Ding, Yuan \& Liang\ \citeyear{Ding2010}), dissipation of MHD turbulence (Quataert \citeyear{Quataert1998}; Blackman \citeyear{Blackman1999}), and dissipation of pressure anisotropy in a collisionless plasma (Sharma et al.\ \citeyear{Sharma2007}). While it is clear that parameters $\alpha$ and $\delta$ are both related with the magnetic field, no consensus on their exact values has been reached so far. This is especially the case for the value of $\delta$ (for a brief summary see Xie \& Yuan\ \citeyear{Xie2012}). From the detailed modelling to the spectrum of various sources, the value of $\delta$ was found to range from $\delta\approx0.5$ in the case of  Sgr A* (Yuan, Quataert \& Narayan \citeyear{YQN2003}) to $\delta\approx0.1$ in the case of some LLAGNs (Yu et al.\ \citeyear{Yu2011}; Liu \& Wu \citeyear{Liu2013}). This could be due to, e.g., the configuration of magnetic field in the fuelling plasma in various objects being different.

\subsection{Jet}
\label{jet}

Details of the internal-shock jet model are described in Yuan, Cui \& Narayan (\citeyear{YCN2005}). The half-opening angle is assumed to be $\theta_{\rm jet}=0.1$, and the bulk Lorentz factor of the jet is set to $\Gamma_{\rm jet} = 1.2$ (or equivalently the bulk velocity is $0.6\ c$) for BHBs and $\Gamma_{\rm jet} = 10$ ($0.99\ c$) for AGNs.

Internal shocks occur in the jet as a result of collisions between shells with different velocities, and a small fraction (typically $1\%$) of the electrons in the jet will be accelerated into a power-law energy distribution with a index $p_{\rm e}$. Because of the efficient radiative cooling of the most energetic electrons, the distribution of the non-thermal electrons will generally become broken power-law, i.e. the steady-state power-law index for the high-energy electrons (responsible for the X-ray radiation) will change into $p_{\rm e}+1$, while the slope for the low-energy electrons (responsible for the radio-IR radiation) will remain $p_{\rm e}$. In our jet model, this change of electron distribution due to radiative cooling is self-consistently calculated. Two additional parameters are $\epsilon_{\rm e}$ and $\epsilon_B$, which describe the fractions of the shock energy that go into power-law electrons and the magnetic fields, respectively. With these non-thermal electrons in magnetic fields, the dominant radiative process is synchrotron radiation, and inverse Compton process is not important.

\subsection{Underlying physics of the $\Gamma$ -- $\lxedd$ correlation}
\label{explanation}

Before providing the quantitative calculations in the next Section, let us describe the main physics. For the very-low X-ray luminosity regime, our model predicts that the X-ray comes from the synchrotron emission of the non-thermal electrons in the jet (Yuan \& Cui \citeyear{Yuan2005}). Because these electrons suffer strong radiative cooling (Heinz \citeyear{Heinz2004}; Yuan \& Cui \citeyear{Yuan2005}; Zdziarski, Lubi\'nski \& Sikora\ \citeyear{Zdziarski2012}), their power-law index of the steady distribution is $p_{\rm e}+1$. Thus, the photon index of the X-ray spectrum will be
\be \Gamma = 1+(p_{\rm e}+1 -1)/2 = 1+p_{\rm e}/2. \label{eq:jet-gamma}\ee
For a typical value of $p_{\rm e}\approx 2.2$ suggested by relativistic shock acceleration theory (Kirk et al.\ \citeyear{Kirk2000}), we have $\Gamma\approx 2.1$. In other words, the jet model {\it predicts} a constant photon index independent of the X-ray luminosity, provided the electron acceleration will lead to the same $p_{\rm e}$ value at different $\dot{M}_{\rm jet}$ (Xie et al.\ \citeyear{Xie2014}).

\begin{figure}
\centering
\includegraphics[width=9cm]{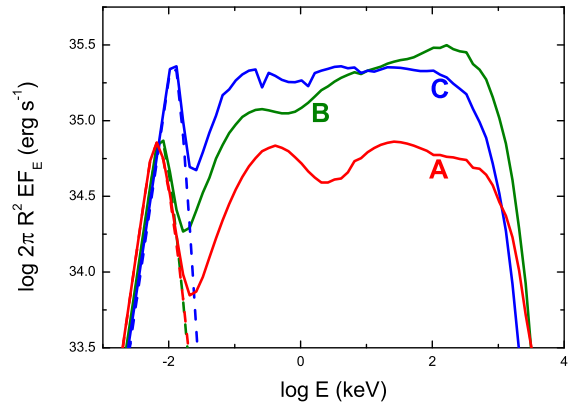}
\caption{The spectral variation of the emission from at $R=5\ R_{\rm S}$ for different accretion rates, namely $\dot{M}=0.4\, \medd$ (Case A, red curves), and $\dot{M}=0.8\, \medd$ (Case B, green curves and Case C, blue curves). The dashed and the solid curves show the synchrotron and Comptonized emission, respectively. Case C has nearly the same dynamical structure (accretion rate, electron density, optical depth and total electron power $L_{\rm hard}$) as Case B, but it has an arbitrarily increased seed photon flux (a factor of $\sim 3$ compared to that of Cases A and B) and its electron temperature is re-calculated from the electron energy balance.}\label{fig:simp}
\end{figure}

For the moderate and high X-ray luminosity branches, the main radiative mechanism, which also dominates the emission in X-ray band, is Compton scattering. The spectrum will be harder (smaller $\Gamma$) provided a larger Compton $y$-parameter, and vise versa. On the other hand, during the investigation of variability of Cyg X-1, Zdziarski et al.\ (\citeyear{zppw02}) considered a simple model of a plasma cloud irradiated by soft photons, and the temperature adjusts itself to the energy balance, i.e. for a given power supplied to the electrons $L_{\rm hard}$, more soft photon power $L_{\rm soft}$ leads to more radiative cooling, lower electron temperature and consequently softer spectrum, and vice versa. Since $L_{\rm hard}$ in reality is not likely a constant, we take a generalized version of this result (see also Done, Gierli\'nski \& Kubota \citeyear{Done2007}), namely the photon index $\Gamma$ positively correlates with $L_{\rm soft}/L_{\rm hard}$. Below we will apply this to understand the observed negative-positive $\Gamma$ -- $\lxedd$ correlation parts. Before that, we emphasise two key factors of our hot accretion flow model. First, unless the density is extremely low, the electrons will emit most of the power they received, i.e. the electrons themselves are radiatively efficient, and the total luminosity can be approximated to be $L_{\rm hard}$ (verified in both ADAF and Type I LHAF, while speculated in two-phase accretion flow. Cf.\ Xie \& Yuan \citeyear{Xie2012}). Second, the origin of seed photons supplied for Compton scattering are different in one-phase hot accretion flows (either ADAFs or Type I LHAFs) and the two-phase accretion flows, i.e. the former is self-absorbed synchrotron from hot gas while the latter is mainly thermal emission from cold clumps.

In the moderate X-ray luminosity regime, a negative $\Gamma$-$\lxedd$ correlation is observed. We interpret it as a hot accretion flow being an ADAF or a Type I LHAF. As the accretion rate increases ($L_{\rm hard}$ also increases), both the electron density and the magnetic strength increase, which lead to an enhanced increase in synchrotron absorption depth. Correspondingly, the seed photon power $L_{\rm soft}$ will increase much slower (compared to $L_{\rm hard}$) due to the enhanced degree of synchrotron self-absorption, i.e. $L_{\rm soft}/L_{\rm hard}$ is negatively correlated with $L_{\rm hard}$. In other words, from the arguments by Zdziarski et al.\ (\citeyear{zppw02}), we will observe a negative correlation between $\Gamma$ and $L_{\rm hard}$ (or equivalently $\lxedd$) for an ADAF/Type I LHAF system.

Then in the bright X-ray luminosity regime, we believe it is the two-phase accretion flow, where it is believed that the fraction of accreting material in cold phase increases with the accretion rate. We thus have the seed photon power $L_{\rm soft}$ increasing faster than the power supplied to the hot electrons $L_{\rm hard}$ (opposite to the case in the ADAF and Type I LHAF), i.e. $L_{\rm soft}/L_{\rm hard}$ is positively correlated with $L_{\rm hard}$. Applying the analysis of Zdziarski et al.\ (\citeyear{zppw02}) we find that the two-phase accretion model has a positive correlation between $\Gamma$ and $\lxedd$, as observed at the high X-ray luminosity branch. We note that the origin of seed photons are different in alternative scenarios, i.e. they can be from the inner small cold disc in the disc -- corona model (Qiao \& Liu \citeyear{Qiao2013a}; Qiao et al.\ \citeyear{Qiao2013b}), or from the outer cold disc in the truncated disc model (Veledina et al.\ \citeyear{Vel2011}; Sobolewska et al.\ \citeyear{Sobolewska2011}; Gardner \& Done \citeyear{Gardner2013}).

In order to quantitatively examine the above analysis on the Comptonized spectrum, we investigate emission at $R=5 R_{\rm S}$ in the BHB case (see Section \ref{results}), with $\delta=0.5$. Instead of an integration in radius, we show $2\pi R^2\ E\ F_E$, where $F_E$ is the radiative flux at the surface. We choose two accretion rates, i.e. $0.4\ \medd$ (Case A) and $0.8\ \medd$ (Case B, a factor of $2$ higher than Case A.). As shown in Fig.\ \ref{fig:simp}, the synchrotron power, $L_{\rm soft}$, slightly changes from Case A to Case B, while the power supplied to the electrons $L_{\rm hard}$ (and also the luminosity of the Comptonized spectrum) increases by a factor of $\sim 2$. Consequently the spectrum hardens, i.e., there is an anti-correlation of $\Gamma$ -- $\lxedd$. In addition, in order to investigate the two-phase accretion flow regime with an increased fraction of cold clumps for increasing $\dot{M}$, we take the dynamical structure (density, velocity, scale height, total electron power $L_{\rm hard}$, etc) of Case B, but arbitrarily increase its $L_{\rm soft}$ by a factor $\sim 3$ in our Case C. We then re-calculate the electron energy balance. Consequently, Case C has the same bolometric luminosity as Case B, but has lower electron temperature. From Fig.\ \ref{fig:simp}, the enhanced $L_{\rm soft}/L_{\rm hard}$ leads to a softer spectrum compared to Case A, i.e., we have a positive $\Gamma$-$\lxedd$ correlation from Case A to Case C.

\section{Calculation Results}
\label{results}

\begin{figure}
\centering
\includegraphics[width=9.cm]{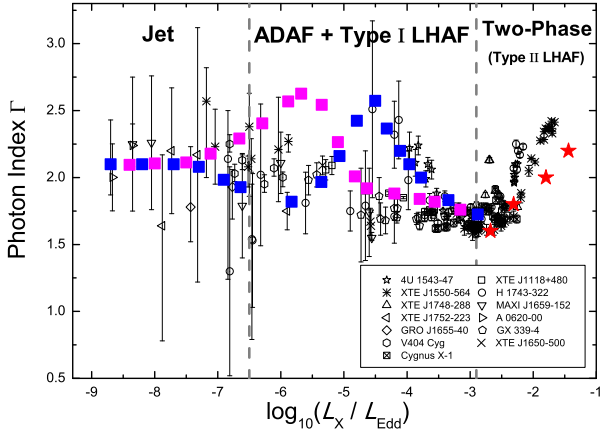}
\caption{The theoretical calculations under the framework of accretion -- jet model for the BHB sample (cf.\ Fig.\ \ref{fig-bhb-sample} for the observational data). The two grey dashed vertical lines separate three regions of the correlation, which we identify theoretically have different X-ray origins as labelled on the top of figure. See text for details. Also shown by the red stars and (blue and pink) squares in the figure are the theoretical modelling results. The parameters are $\alpha=0.3$, $s=0.3$, $\beta=9$, and $p_{\rm e} = 2.2$. The values of $\delta$ are $\delta=0.5$, $0.1$, and $0.5$ for the blue and pink solid squares, and the red stars, respectively. }\label{fig-bhb-theory}
\end{figure}

\subsection{General results}
\label{general-results}

We set $M_{\rm BH}=10\, \msun$ and $10^8\, \msun$ for BHBs and AGNs, respectively, $\alpha=0.3$, the outflow parameter $s=0.3$ (Yuan, Wu \& Bu \citeyear{YWB2012}), and the truncated radius $R_{\rm tr} = 10^3 R_{\rm S}$. In the actual flow, we expect the value of $R_{\rm tr}$ to be smaller at relatively large accretion rates (Yuan \& Narayan \citeyear{Yuan2014}). The Compton scattering of emission from outer cold disc is neglected in our calculation of the X-ray spectrum. This is a good approximation for ADAF and Type I LHAF, where $R_{\rm tr}$ is relatively large and the emission and reflection angle of the outer thin disc is small. For two-phase accretion flow, $R_{\rm tr}$ is likely small, and the cold photons from outer thin disc will be an important seed photon source for the hot accretion flow. However, its impact can be absorbed in the $y$ parameter, which we assume arbitrarily, and which also takes into account the effect of cold clumps. For BHBs, we assume $\beta=9$ and consider two cases of $\delta$, $\delta=0.1, 0.5$, while for AGNs, we assume that $\beta$ and $\delta$ are different for various objects and in the ranges of $1\leq \beta\leq 9$ and $0.1 \leq \delta \leq 0.5$. For the two-phase accretion flow, since our current calculations of the dynamics and radiation are crude, we choose only one parameter set, namely, $\beta=9$ and $\delta=0.5$.

For the jet model, we set the viewing angle $\theta_{\rm obs} = 60\degr$, $\epsilon_{\rm e} = 0.2$, $\epsilon_B = 0.02$, and $p_{\rm e} = 2.2$. We then use Fig.\ 2 in Yuan \& Cui (\citeyear{Yuan2005}) to determine the ratio of the mass loss rate in the jet to the mass accretion rate in the innermost region of the accretion flow for a given model. The ratio is obtained by modelling the observed radio/X-ray correlation of BHBs and AGNs\footnote{That figure is only for cases of $\delta=0.01$ and $0.5$. We now also consider cases of other values of $\delta$.}. Note that since we assumed different values of $\Gamma_{\rm jet}$ for BHBs and AGNs, the corresponding Doppler factors are different, $\simeq 1.2$ and $\simeq 0.2$, respectively. This results in a strong de-boosting of the observed emission in the AGN case, which we compensate for (admittedly arbitrarily) by an increase of $\dot{M}_{\rm jet}$.

\begin{figure}
\centering
\includegraphics[width=9.cm]{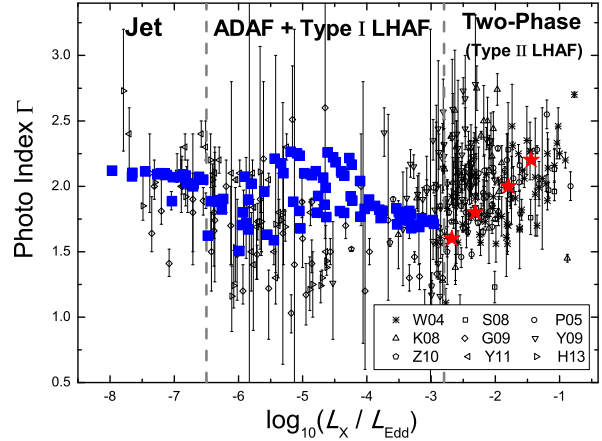}
\caption{The theoretical calculations under the framework of accretion -- jet model for the AGN sample (cf.\ Fig.\ \ref{fig-agn-sample} for the observational data). The two grey dashed vertical lines separate three regions of the correlation, which we identify theoretically have different X-ray origins, as labelled on the top of the figure. Also shown by the squares and stars are the theoretical modelling results. The parameters are $\alpha=0.3$, $s=0.3$ and $p_{\rm e} = 2.2$. The other parameter are $1 \leq \beta \leq 9$ and $0.1 \leq \delta \leq 0.5$ for the blue squares, and $\beta = 9$ and $\delta = 0.5$ for the red stars.}\label{fig-agn-theory}
\end{figure}

Our numerical results are shown in Figs.\ \ref{fig-bhb-theory} and \ref{fig-agn-theory} for BHBs and AGNs respectively. The vertical dashed lines denote the observed luminosity boundaries between the three regimes (see Section \ref{sample}). The labels in each figure, namely, ``Jet", ``ADAF+Type I LHAF'', and ``Two-Phase'', indicate the respective dominate contributor to the X-ray radiation. We can see that the model explains the observations reasonably well on average.

However, we can see from the figures that when $10^{-6.5}\la  l_{\rm X}\la  10^{-4.5}$, there seems to exist a positive correlation, or a ``bump'' in the calculated correlation curve. This is more clearly seen in Fig.\ \ref{fig-bhb-theory}.  The reason for such a ``bump'' is as follows. At such low accretion rates, the scattering optical depth of the accretion flow is very low. In this case, the resulted spectrum of Compton scattering is not a power-law form, but a curve with scattering bumps (cf.\ Fig. 1 in Yuan \& Narayan \citeyear{Yuan2014}). In this case, the simple analysis in Section \ref{explanation} fails. The positive correlation in the range of $10^{-6.5}\la  l_{\rm X}\la  10^{-4.5}$ corresponds to the rising part of the bump. Due to the difference of black hole masses, the 2--10 keV radiation is dominated by the first and second Compton scattering bumps for BHBs and AGNs respectively. This prediction is still difficult to examine by current observations, mainly because of the large error bar in the value of $\Gamma$. This is also the case when we compare the theoretical result with the observation of individual sources (refer to Fig.\ \ref{fig-two-bhbs} and Section \ref{individualBHB}). But there seems to be some tentative sign of such a bump in the correlation in both Figs. \ref{fig-bhb-sample} and \ref{fig-agn-sample}. Further investigations are highly desirable.
{\bf In our hot accretion flow model, we only consider thermal electrons. If we consider the possible existence of non-thermal electrons in the hot accretion flow, as proposed by, e.g., Veledina et al.\  (\citeyear{Vel2011} and references therein), the scattering bump will become less obvious or even disappear. Consequently, the ``bump'' in the $\Gamma$ vs.\ $l_{\rm X}$ correlation will also become weaker or  disappear. Better observational data in future may put some constraint on the existence of non-thermal electrons in the hot accretion flow.}

\subsection{Turning point at $l_{\rm x} \sim 10^{-3}$ -- location of the transition in accretion mode}
\label{point}

As stated in the Introduction, there exist a turning point in the $\Gamma$ -- $\lxedd$ correlation, which, from the sample we combined (see Section \ref{sample}, and Figs. \ref{fig-bhb-theory} \& \ref{fig-agn-theory}), locates at $\lxedd \sim 10^{-3}$ (e.g., Yuan et al.\ \citeyear{Yuan2007}; Constantin et al.\ \citeyear{Constantin2009}; Veledina et al.\ \citeyear{Vel2011}; Trichas et al.\ \citeyear{Trichas2013}). If we take the correction factor between the 2--10 keV X-ray luminosity and the bolometric luminosity $L_{\rm bol}$ to be $\sim 1/16$ (Ho \citeyear{Ho2008}), then the turning point is at  $L_{\rm bol}/L_{\rm Edd}\sim (1-2)\%$.

What is the physical origin of this turning point? In our hot accretion -- jet model, it corresponds to the change of accretion mode between Type I LHAF (an one-phase accretion flow) and two-phase accretion flow. The corresponding luminosity of the turning point predicted by theoretical calculations is also consistent with the observed value. In the evaporated-corona scenario, the turning point corresponds to the formation of an inner cold disc from the original hot corona (Liu, Done \& Taam\ \citeyear{Liu2011}; Qiao \& Liu \citeyear{Qiao2013a}).

\subsection{The correlation of individual BHBs}
\label{individualBHB}

\begin{figure}
  \centering
  \includegraphics[width=90mm]{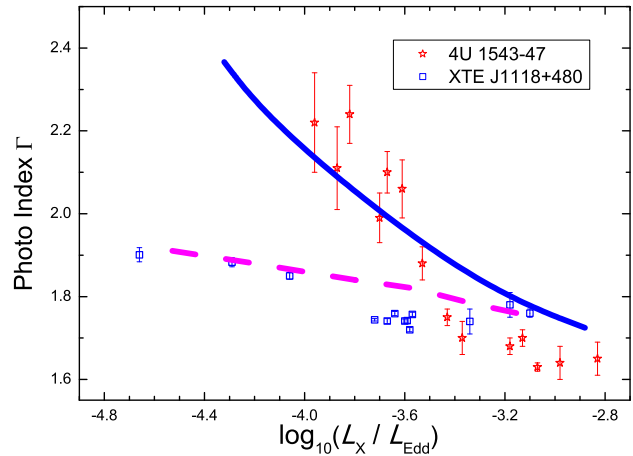}
  \caption{The $\Gamma$ -- $\lxedd$ correlation for two individual BHBs (which is the zoom in of Fig.\ \ref{fig-bhb-sample} in the luminosity range $10^{-4.8}\la  l_{\rm X} \la  10^{-2.8}$). The correlation in each individual source is much tighter, and the slope of the correlation varies for different sources. The theoretical modeling results are also shown in the figure by the blue solid ($\delta=0.5$) and pink dashed ($\delta=0.1$) lines. See Section \ref{individualBHB} for details. }\label{fig-two-bhbs}
\end{figure}

In Figs.\ \ref{fig-bhb-sample} and \ref{fig-agn-sample}, the $\Gamma$ vs.\ $l_{\rm X}$ correlations for both the BHB and AGN samples show large scatter. We speculate that the scatter takes place because some parameters of accretion flow are different from source to source. This is confirmed by Figs.\ \ref{fig-two-bhbs} and \ref{fig-ngc7213} in the sense that the correlation becomes much tighter for individual sources, but with their own correlation slope. Through detailed calculations, we identify in this and next sections that one such parameter of hot accretion flow is $\delta$. The value of $\delta$ should remain roughly constant when the accretion rate changes during their evolution while different for different sources. We note that it will be too ambiguous for any quantitative investigation on the positive correlation regime, because our two-phase accretion model is far from mature, i.e. the correlation slope is sensitive to the assumed $y$--$\dot{M}$ relationship. We thus focus on the negative correlation part.

Fig.\ \ref{fig-two-bhbs} shows the correlation for two BHBs, with $l_{\rm X}$ lying in the range of $10^{-4.5}$--$10^{-3}$. Theoretical modelling with $\beta=9$ but $\delta=0.1$ (pink dashed) and $0.5$ (blue solid) are also shown there. We can see that for a larger $\delta$ the $\Gamma$ -- $\lxedd$ correlation is steeper. From this plot, we speculate that the $\delta$ value in 4U 1543-47 is larger than that in XTE J1118+480. The dependence of the slope of the correlation on $\delta$ can be understood as follows. Firstly, at the low luminosity ADAF end, for a given X-ray luminosity, since the radiative efficiency of ADAF is sensitive to $\delta$ (Xie \& Yuan \citeyear{Xie2012}), systems with larger $\delta$ requires a smaller accretion rate. This in turn results in a softer spectrum since optical depth is small. Secondly, at the luminous Type I LHAF end, due to the strong compensation of the Coulomb interaction energy exchange, the radiative efficiency of Type I LHAF is insensitive to $\delta$ (Xie \& Yuan \citeyear{Xie2012}), which implies that the value of $\Gamma$ is almost the same for different sources with different $\delta$. As discussed in Section \ref{hotaccretion}, the difference of $\delta$ may be a result of the differences in the configuration and/or strength of magnetic field in the accretion flow.

Hysteresis behaviour could also be a reason for the different correlation slope. There is no cold disc initially during the rising, hard-state phase of an outburst, while during the declining phase of the outbursts the transition from the soft to the hard state begins with the cold disc extending all the way to the innermost stable circular orbit. This makes a clear difference, i.e. the same $\Gamma$ may appear at X-ray luminosities that differ by 1--2 orders of magnitude.

\subsection{The correlation of individual AGNs}
\label{individualAGN}

Because of the short evolutionary timescale, it is relatively easy to obtain the $\Gamma$ -- $\lxedd$ correlation of an individual BHB (cf.\ Fig.\ \ref{fig-bhb-sample} \& \ref{fig-two-bhbs}). In contrast, it is much more difficult to obtain such correlation of an individual AGN. Only a handful AGNs have long-term observations with adequate coverage in $L_{\rm X}$ for the correlation investigation here. We will discuss them in this subsection.

In the high luminosity branch, Sobolewska \& Papadakis (\citeyear{Sobolewska2009}) investigated the long time-scale X-ray spectral variability of 10 nearby bright AGNs by \rxte\ over periods of $7-11$ yrs. Based on their X-ray luminosity ratio ($\lxedd$), these sources are expected to lie in the two-phase accretion regime thus a positive correlation is expected. The observed positive $\Gamma$ -- $\lxedd$ correlation for these sources confirms our expectation. Additional bright sources with positive correlation are reported in Zdziarski \& Grandi\ (\citeyear{ZG2001}) and Zdziarski et al.\ (\citeyear{Zdziarski2003}).

\begin{figure}
\center
\includegraphics[width=90mm]{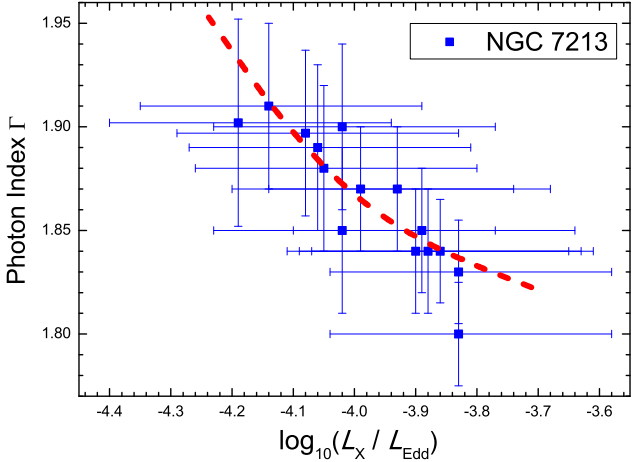}
\caption{The $\Gamma$ -- $\lxedd$ correlation for an individual low-luminosity AGN, NGC~7213. The observational data are from Emmanoulopoulos et al.\ (2012). The red dashed line shows the theoretical modelling result, with $\beta = 1$ and $\delta=0.45$.}\label{fig-ngc7213}
\end{figure}

In the low luminosity branch, to our knowledge currently there is only one source that has the correlation data, which comes from the long-term \rxte\ observations of NGC~7213 (Emmanoulopoulos et al.\ \citeyear{Emmanoulopoulos2012}). The multi-waveband spectral energy distribution of this source, especially the absence of the big-blue-bump, also favors the truncated disc model (Emmanoulopoulos et al.\ \citeyear{Emmanoulopoulos2012}). Fig.\ \ref{fig-ngc7213} shows the correlation of this source. Although the X-ray luminosity only spans by a factor $\sim 4$--5, it clearly shows a ``harder when brighter'' spectral behaviour, consistent with our model. The mass of the black hole is $9.6^{+6.1}_{-4.1}\times 10^7\msun$ (Bell et al.\ \citeyear{Bell2011}).  In our numerical calculation, we adopt the parameters \mbh=$10^8\ \msun$, $\alpha=0.3, \beta = 1$ and $\delta=0.45$. As shown by the red dashed line in Fig.\ \ref{fig-ngc7213}, it quantitatively agrees very well with the observations.

\section{Summary and Discussions}
\label{summary}

\subsection{Summary}
\begin{figure}
\center
\includegraphics[width=80mm]{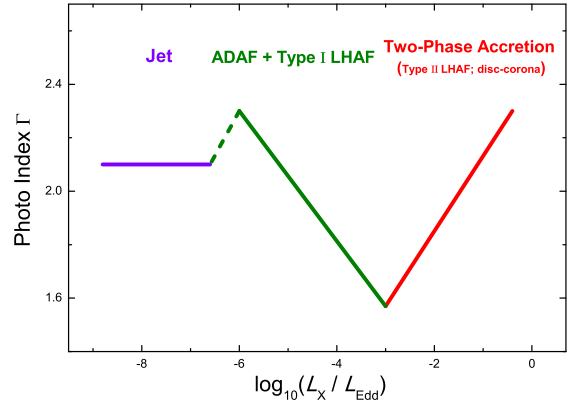}
\caption{Schematic diagram of the 2--10 keV X-ray photon index $\Gamma$ versus the 2--10 keV luminosity ratio $\lxedd$. Labels with the same color mark the key component in the (hot) accretion -- jet model that is responsible for the correlation.}\label{fig-schematic-diagram}
\end{figure}

In this work, we have combined the data of BHBs and AGNs from literature to construct the largest sample so far to investigate the correlation between the X-ray photon index $\Gamma$ and the 2--10 keV X-ray luminosity (in Eddington unit; $\lxedd$). The $\lxedd$ in our sample ranges from $\sim10^{-9}$ to $\sim10^{-1}$, much larger than all previous investigations. We have confirmed the existence of a correlation between these two quantities, although the correlation has substantial intrinsic scatter (see Figs.\ \ref{fig-bhb-sample} and \ref{fig-agn-sample} for the correlation in BHB and AGN samples, respectively). More specifically, the correlation consists of three parts, which can be naturally interpreted in the framework of a coupled hot accretion flow plus jet model. Here the hot accretion flow includes the ``purely'' hot flow with low accretion rates and two-phase accretion flow with high accretion rates. As summarized schematically in Fig.\ \ref{fig-schematic-diagram}, our main results are as follows.

\begin{itemize}
  \item {\bf Jet} In the extremely low luminosity range, i.e., $l_{\rm X} \la  10^{-6.5}$, the value of $\Gamma$ is constant. This is because our model predicts that in this luminosity regime the X-ray emission should be dominated by jet. The photon index is thus determined by the energy distribution of the power-law electrons in the jet, a distribution which is roughly the same for different sources and different $l_{\rm X}$. The observed $\Gamma\approx 2.1$ is consistent with synchrotron emission from radiatively cooled power-law electrons, with an injected energy distribution $p_{\rm e} = 2.2$, a value suggested from relativistic shock acceleration theory.
  \item {\bf ADAF+Type I LHAF} In the moderate luminosity range, i.e., $10^{-6.5}\la  l_{\rm X} \la  10^{-3}$, a negative correlation is observed. In our model the X-ray radiation in this luminosity regime comes from the one-phase hot accretion flow. The self-absorbed synchrotron emission provides the seed photons for the Compton scattering. We note that a bump exists in the predicted correlation, mainly because of the appearance of ``Compton bumps'' in the spectrum when the accretion rate is small. Only a tentative sign of this bump exists in the observational data and better data is required to examine this prediction. We note that these bumps may become less obvious for the presence of non-thermal electrons in hot accretion flows.
  \item {\bf Two-Phase Accretion Flow} In the high luminosity range, i.e., $l_{\rm X} \ga  10^{-3}$, a positive correlation is observed. In our picture the accretion flow in this regime will be in two-phase. The Compton up-scattering of the optical/ultraviolet radiation from the cold clumps will highly cool the electrons in the hot phase. If our assumption of the anti-relationship between Compton $y$ parameter and $\dot{M}$ is reasonable (cf.\ Sections\ \ref{hotaccretion} \&\ref{explanation}), a positive relationship will then be expected.
  \item The correlation has a large scatter. We argue that one important reason is that the model parameter, likely $\delta$, is different for different sources. This is supported by the fact that the correlation of each individual sources is much tighter, while the correlation slope is different from source to source. We have confirmed this speculation by interpreting the correlation of individual sources based on this idea.
\end{itemize}

\subsection{Caveats}
\label{sec:caveats}

Some simplifications are adopted in our calculations. One is the neglect of the non-thermal electrons in the hot accretion flow, whose existence is still difficult to constrain (for the impact of non-thermal electrons, see e.g.  Malzac \& Belmont\ \citeyear{MB2009}; Poutanen \& Vurm\ \citeyear{Poutanen2009}; Veledina et al.\ \citeyear{Vel2011}; Poutanen \& Veledina \citeyear{Poutanen2014}; Nied{\'z}wiecki, Xie \& Stepnik \citeyear{Nied2014}; Nied{\'z}wiecki, Stepnik \& Xie \citeyear{Nied2015}). In addition, our model is non-relativistic, and a pseudo-Newtonian potential is adopted. Moreover, global Compton scattering, which is important for bright sources (Yuan, Xie \& Ostriker \citeyear{YXO2009}; Xie et al.\ \citeyear{Xie2010}; Nied{\'z}wiecki, Xie \& Zdziarski \citeyear{Nied2012}, and references therein), is not taken into account for simplicity. Using a full general relativistic (in both dynamics and radiation) ADAF model with global Compton scattering, the negative $\Gamma$ -- $\lxedd$ correlation is investigated extensively by Nied{\'z}wiecki, Xie \& Stepnik (\citeyear{Nied2014}).

Our AGN sample excludes those high-redshift ($z\ga1.0$) sources whose black hole masses are constrained through C{\sc iv}-based reverberation mapping method. Most of these sources have X-ray luminosities larger than $10^{-3}\ L_{\rm Edd}$ while a negative correlation is observed (Kelly et al.\ \citeyear{Kelly2008}; but see Risaliti, Young \& Elvis\ \citeyear{Risaliti2009} for a smaller C{\sc iv}-determined-$M_{\rm BH}$ AGN sample). This is quite different from other sources investigated here, whose black hole masses are constrained through \hb emission line (note that sources whose black hole masses are constrained through Mg{\sc ii} emission line show similar results as those through \hb line). It is unclear whether this is because the estimation of the black hole mass by C{\sc iv} line is problematic.

In this work, we only focus on sub-Eddington sources, i.e. $L_{\rm bol} \la L_{\rm Edd}$. It is interesting to note that a positive correlation similar to that shown by equation \ref{eq:pos-corr-agn} seems to exist for super-Eddington AGNs as well (e.g., Bian \citeyear{Bian2005}). This may imply that the disc -- corona model may also work for those sources. Detailed discussions are beyond the scope of the present paper.

\subsection{Two-phase accretion flow: clumpy and the disc-corona models}

A widely discussed scenario in the literature for the origin of X-ray emission of luminous sources is the ``disc-corona'' model, i.e., a hot corona sandwiching a thin disc (e.g., Liang \& Price \citeyear{Liang1977}; Galeev, Rosner \& Vaiana \citeyear{Galeev1979}; Haardt \& Maraschi \citeyear{Haardt1991}; Merloni \& Fabian \citeyear{Merloni2001}; Liu, Mineshige \& Ohsuga \citeyear{Liu2003}; Cao \citeyear{Cao2009}; Schnittman, Krolik \& Noble \citeyear{Schnittman2013}). What is the relation between this scenario and the two-phase accretion scenario adopted in the present work? In our understanding, the two-phase accretion scenario includes this disc-corona model but is more general in the following sense. With the increase of mass accretion rate, as we have mentioned in Section \ref{hotaccretion}, some cold dense clumps will be formed, embedding in the hot phase. When the mass accretion rate is not very large, the cold clumps are in the state of individual clouds and are suspending in the hot gas. We call this accretion flow ``clumpy''. With the further increase of accretion rate, more and more clumps are condensed out and they will gradually merge together and settle down to form a thin disc in the equatorial plane, with hot plasma sandwiching the thin disc. This is the usual disc-corona model.

In our model, the positive correlation (i.e., the red solid line in Fig.\ \ref{fig-schematic-diagram}) is explained by the two-phase accretion flow model (Type-II LHAF). Another possibility proposed in previous works is the disc-corona model, i.e., the two-phase accretion flow with high accretion rates (e.g., Cao \& Wang \citeyear{Cao2014}). We argue that the disc-corona model may not be able to explain the correlation when $10^{-3} \la  l_{\rm X} \la  10^{-1.5}$, i.e., the low-luminosity part of the red line close to the turning point, but only the high-luminosity part at $l_{\rm X}\ga 10^{-1.5}$. The photon index at the low-luminosity part is small, i.e., the spectrum is hard. But for the disc-corona model, since there are abundant soft photons emitted from the thin disc, the predicted X-ray spectrum is usually very soft, i.e. $\Gamma \ga 2$ (e.g. Haardt \& Maraschi \citeyear{Haardt1991}; Haardt, Maraschi \& Ghisellini\ \citeyear{HMG94}; Stern et al.\ \citeyear{S95} and recently Cao \citeyear{Cao2009})\footnote{We note that the spectra in Schnittman et al.\ (\citeyear{Schnittman2013}) are actually very soft ($\Gamma\sim3$) in the 2--10\ ${\rm keV}$ band, cf.\ their Fig. 12.}. On the other hand, we see from Figs. \ref{fig-bhb-sample} and \ref{fig-agn-sample} that there are many sources with $\Gamma\la 2.0$ in this positive correlation branch. As we have shown in Section \ref{results}, the ``clumpy'' two-phase accretion flow model adopted in the present work can well explain this part. Another argument against the disc-corona model as the interpretation of the positive correlation will be described in the next subsection (Section \ref{luminous-AGNs}).

\subsection{Accretion model for the X-ray emission in some luminous AGNs}
\label{luminous-AGNs}

In addition to the problem of the spectral fitting, another argument against the disc-corona model as an explanation of the overall positive correlation comes from the geometry of the accretion flow. For BHBs, all of the data points in the positive correlation branch (cf.\ Section \ref{BHBs} and Fig. \ref{fig-bhb-sample}) come from the intermediate (and bright hard) state. As we know, many observational and theoretical works have shown that for the hard and intermediate states the thin disc must be truncated at a certain radius (for BHBs, see Tamura et al.\ \citeyear{Tamura2012}; Yamada et al.\ \citeyear{Yamada2013}; Hori et al.\ \citeyear{Hori2014}; see Yuan \& Narayan \citeyear{Yuan2014} for more evidence). This precludes the disc-corona model as the sole model of the positive correlation of BHB sample since in this model the thin disc extends to the ISCO without truncation.

For the AGN sample, the positive correlation consists of data from luminous AGNs such as quasars. Compared to the BHB sample, we can see from Fig.\ \ref{fig-agn-sample} that the ``turning point'' is also at $l_{\rm x}\sim 10^{-3}$, but the luminosity extends to higher values, $l_{\rm x}\ga 10^{-1.5}$. So we speculate that the disc-corona model may be the correct model to interpret the part of correlation beyond $10^{-1.5}$, where the spectrum is rather soft with photon index $\Gamma\ga 2$. We also note that, when $l_{\rm x}\ga 10^{-1.5}$, the BHBs will generally be in the soft state, a situation disc-corona model likely operates. For observations in the soft (and brighter) state, a positive correlation in $\Gamma$ -- $\lxedd$ is also found, like the case of bright AGNs.

In summary, the two-phase accretion flow model, including the clumpy hot accretion flow model at relatively lower accretion rates and the disc-corona model at higher accretion rates, can explain the luminous AGNs. This is consistent with the results obtained in Yuan \& Zdziarski (\citeyear{Yuan2004}). The clumpy and the disc-corona models are responsible for the X-ray origin of AGNs with $l_{\rm x}\la 10^{-1.5}$ and $l_{\rm x}\ga 10^{-1.5}$ respectively.

\section*{Acknowledgments}
We thank the anonymous referee for constructive suggestions, and Qingwen Wu, Zhaozhou Li, and Xiaobo Dong for useful discussions. This work was in part supported by the Strategic Priority Research Program ``The Emergence of Cosmological Structures'' of the Chinese Academy of Sciences (Grant XDB09000000), the National Basic Research Program of China (973 Program, grant 2014CB845800), and the Natural Science Foundation of China (grants 11203057, 11103061, 11133005 and 11121062), and was also supported in part by the Polish NCN grants 2012/04/M/ST9/00780 and 2013/10/M/ST9/00729.

\label{lastpage}

\end{document}